\documentclass{article}
\usepackage{cite}
\usepackage{graphicx}
\usepackage{dcolumn}

\begin{document}

\date{}
\title{Dimensionless equations in non-relativistic quantum mechanics}
\author{Francisco M. Fern\'{a}ndez \thanks{%
E-mail: fernande@quimica.unlp.edu.ar} \\
INIFTA, Divisi\'on Qu\'imica Te\'orica\\
Blvd. 113 S/N, Sucursal 4, Casilla de Correo 16\\
1900 La Plata, Argentina}
\maketitle

\begin{abstract}
We discuss the numerous advantages of using dimensionless equations in
non-relativistic quantum mechanics. Dimensionless equations are considerably
simpler and reveal the number of relevant parameters in the models. They are
less prone to round-off errors when applying numerical methods because all
the quantities are of the other of unity. A dimensionless equation
facilitates the application of perturbation theory and provides a glimpse of
the sort of solution we are going to obtain beforehand.
\end{abstract}

\section{Introduction}

\label{sec:intro}

Solving the Schr\"{o}dinger equation in non-relativistic quantum mechanics
is greatly facilitated if we first convert that equation into a
dimensionless form. The reason is that fundamental constants like $\hbar $
(Plank constant divided $2\pi $) the electronic mass $m_{e}$ and charge $e$,
as well as other model parameters, are removed leaving a much simpler
equation\cite{F01}. The algebraic manipulation of the dimensionless equation
is considerably less laborious and its numerical treatment, if it is not
exactly solvable, exhibits less round-off errors after getting rid of such
small numbers.

It is a common practice, though in our opinion quite undesirable, to state
that ``we choose units so that $\hbar =m=e=c=1$'' or similar expressions\cite
{AD02}. This is specially so in the case of pedagogical papers where one
should teach the students to carry out the procedure of making dimensionless
equations in detail\cite{F03}. For this reason, in this paper we show how to
derive dimensionless equations and discuss the advantages of the approach as
well as valuable information about the physical result we are looking for.

In section~\ref{sec:one-dim} we discuss several one-dimensional examples, in
section~\ref{sec:Atoms_molecules} we focus on atoms and molecules, in
section~\ref{sec:PT} we outline the utility of dimensionless equations in
the application of perturbation theory and in section~\ref{sec:conclusions}
we summarize the main results and draw conclusions.

\section{One-dimensional models}

\label{sec:one-dim}

In order to illustrate how to convert quantum-mechanical equations into
dimensionless differential equations we begin with some simple
one-dimensional models in non-relativistic quantum mechanics. For
simplicity, we first focus on time-independent problems. The Hamiltonian
operator for a particle of mass $m$ in a potential $V(x)$ in the coordinate
representation is given by
\begin{equation}
H=-\frac{\hbar ^{2}}{2m}\frac{d^{2}}{dx^{2}}+V(x),  \label{eq:H_1D_gen}
\end{equation}
where $-\infty <x<\infty $. The strategy is simple: we first define a
dimensionless coordinate $\tilde{x}\equiv x/L$, where $L$ is a unit of
length that we choose conveniently for each problem. If we take into account
that $d/dx=\left( d\tilde{x}/dx\right) d/d\tilde{x}=L^{-1}d/d\tilde{x}$ we
conclude that $d^{2}/dx^{2}=L^{-2}d/d\tilde{x}^{2}$. Second, we define the
dimensionless Hamiltonian operator $\tilde{H}$ as
\begin{equation}
\tilde{H}=\frac{mL^{2}}{\hbar ^{2}}H=-\frac{1}{2}\frac{d^{2}}{d\tilde{x}^{2}}%
+\frac{mL^{2}}{\hbar ^{2}}V(L\tilde{x}).  \label{eq:H_1D_gen_dim}
\end{equation}
Therefore, if $\psi $ is an eigenfunction of $H$ with eigenvalue $E$ ($H\psi
=E\psi $) then the corresponding eigenvalue $\tilde{E}$ of $\tilde{H}$ is
the dimensionless energy and both eigenvalues are related by
\begin{equation}
E=\frac{\hbar ^{2}}{mL^{2}}\tilde{E},  \label{eq:E_dim}
\end{equation}
where $\hbar ^{2}/\left( mL^{2}\right) $ is the unit of energy.

As a first example we choose a particle of mass $m$ in an impenetrable box
of length $L$
\begin{eqnarray}
H\psi &=&-\frac{\hbar ^{2}}{2m}\frac{d^{2}}{dx^{2}}\psi (x)=E\psi (x),
\nonumber \\
\psi (0) &=&\psi (L)=0.  \label{eq:particle_in _box}
\end{eqnarray}
In this case the unit of length is given by the box length and we define $%
\tilde{\psi}\left( \tilde{x}\right) \equiv \psi \left( L\tilde{x}\right) $
(the normalization factor is irrelevant for present discussion) that is a
solution to
\begin{equation}
-\frac{1}{2}\frac{d^{2}}{d\tilde{x}^{2}}\tilde{\psi}\left( \tilde{x}\right) =%
\tilde{E}\tilde{\psi}\left( \tilde{x}\right) ,
\label{eq:particle_in_box_dim}
\end{equation}
and satisfies the boundary conditions $\tilde{\psi}\left( 0\right) =\tilde{%
\psi}\left( 1\right) =0$. The eigenvalues of this equation are
\begin{equation}
\tilde{E}_{n}=\frac{n^{2}\pi ^{2}}{2},\;n=1,2,\ldots ,
\label{eq:E_n_particle_in_box_dim}
\end{equation}
so that equations (\ref{eq:E_n_particle_in_box_dim}) and (\ref{eq:E_dim})
yield the well known energies of the particle in a box\cite{M76}. It is
clear that when we make the quantum-mechanical eigenvalue equation
dimensionless we can predict the dependence of the energies on the model
parameters and physical constants ($\hbar $, $m$ and $L$ in the present
case) without solving the equation. Besides, equation (\ref
{eq:particle_in_box_dim}) makes the statement ``we choose units so that $%
\hbar =m=L=1$'' self-evident (specially for pedagogical purposes). In
addition to it, we also predict that the eigenfunctions for the particle in
a box will depend on the variable $\tilde{x}\equiv x/L$ and this is actually
so as shown by\cite{M76}
\begin{equation}
\psi _{n}=\sqrt{\frac{2}{L}}\sin \left( \frac{n\pi x}{L}\right) .
\end{equation}
It is always convenient to tell the students that the arguments of functions
like $\sin (z)$, $\ln (z)$, $\exp (z)$, etc should be dimensionless and that
a result that does not follow this rule is wrong. In our opinion, the
dimensionless energy $\tilde{E}$ is more convenient than $\epsilon \equiv
2mE/\hbar ^{2}$\cite{M76} that exhibits units of length$^{-2}$.

The example above may look unimpressive and we will discuss some others
below. Another simple one is the harmonic oscillator with potential
\begin{equation}
V(x)=\frac{k}{2}x^{2},  \label{eq:V_HO}
\end{equation}
where $k>0$ is the force constant. In this case we have
\begin{equation}
\frac{mL^{2}}{\hbar ^{2}}V(L\tilde{x})=\frac{mL^{4}k}{\hbar ^{2}}\frac{%
\tilde{x}^{2}}{2}.
\end{equation}
Therefore, if we choose
\begin{equation}
L\equiv \left( \frac{\hbar ^{2}}{mk}\right) ^{1/4},  \label{eq:L_HO}
\end{equation}
then the dimensionless Hamiltonian operator will be
\begin{equation}
\tilde{H}=-\frac{1}{2}\frac{d^{2}}{d\tilde{x}^{2}}+\frac{1}{2}\tilde{x}^{2},
\label{eq_H_HO_dim}
\end{equation}
and the unit of energy results to be
\begin{equation}
\frac{\hbar ^{2}}{mL^{2}}=\hbar \omega ,\;\omega =\sqrt{\frac{k}{m}}.
\label{eq:unit_E_HO}
\end{equation}
We realize that the energies of the harmonic oscillator are of the form $%
E_{n}=\hbar \omega \tilde{E}_{n}$, where $\tilde{E}_{n}$ is dimensionless.
If we solve the eigenvalue equation we obtain the allowed energies $%
E_{n}=\hbar \omega \left( n+1/2\right) $, $n=0,1,\ldots $ that agree with
the previous equation because $\tilde{E}_{n}=\left( n+1/2\right) $ are the
eigenvalues of the dimensionless Hamiltonian operator (\ref{eq_H_HO_dim})%
\cite{M76}. Once again, we realize the form of the energies in terms of the
model parameters and physical constants $\hbar $, $m$ and $k$ without
solving the eigenvalue equation. Besides, we also know that the states of
the harmonic oscillator will be functions of $\tilde{x}$. It is well known
that in this case $\psi _{n}(x)=N_{n}H_{n}\left( \tilde{x}\right) \exp
\left( -\tilde{x}^{2}/2\right) $, where $H_{n}\left( \tilde{x}\right) $ is
an Hermite polynomial and $N_{n}$ a normalization factor\cite{M76}. Notice
that the definition (\ref{eq:L_HO}) makes $\frac{\hbar ^{2}}{2mL^{2}}$
(which resembles the kinetic energy) equal to $\frac{kL^{2}}{2}$ (which
resembles the potential energy).

In some cases the potential-energy function is defined in terms of a length
parameter, for example:
\begin{equation}
V(x)=V_{0}f\left( \frac{x}{a}\right) .  \label{eq:V=V0_f}
\end{equation}
In this case
\begin{equation}
\frac{mL^{2}}{\hbar ^{2}}V\left( L\tilde{x}\right) =\frac{mL^{2}}{\hbar ^{2}}%
V_{0}f\left( \frac{L\tilde{x}}{a}\right) ,
\end{equation}
and we have two possibilities. If we choose $L\equiv a$ we have
\begin{eqnarray}
\tilde{H} &=&\frac{ma^{2}}{\hbar ^{2}}H=-\frac{1}{2}\frac{d^{2}}{d\tilde{x}%
^{2}}+\lambda f\left( \tilde{x}\right) ,\;\lambda =\frac{ma^{2}}{\hbar ^{2}}%
V_{0},  \nonumber \\
\frac{ma^{2}}{\hbar ^{2}}E &=&\lambda \tilde{E},\;\tilde{E}=\frac{E}{V_{0}}.
\label{eq:H_f_dim_1}
\end{eqnarray}
If, on the other hand, we choose
\begin{equation}
L\equiv \frac{\hbar }{\sqrt{mV_{0}}},
\end{equation}
then
\begin{equation}
\tilde{H}=\frac{1}{V_{0}}H=-\frac{1}{2}\frac{d^{2}}{d\tilde{x}^{2}}+f\left(
\frac{\tilde{x}}{\sqrt{\lambda }}\right) .  \label{eq:H_f_dim_2}
\end{equation}
Although the dimensionless Hamiltonians (\ref{eq:H_f_dim_1}) and (\ref
{eq:H_f_dim_2}) are different and have different eigenvalues in both cases $%
E=V_{0}\tilde{E}\left( \lambda \right) $. The choice of $L$ depends on what
we are planning to do with the resulting dimensionless equation. The
examples in section\ref{sec:PT} will show the utility of this aparent
arbitrariness.

Up to now we have been tacitly assuming that our interest was the
calculation of the bound states supported by the potential. Suppose that we
are interested in the calculation of the scattering states for the potential
(\ref{eq:V=V0_f}). In this case the present analysis tells us that the
transmission $T$ and reflection $R$ coefficients can be expressed in terms
of only two quantities: $\lambda $ and $\tilde{E}$. For instance, consider
the textbook example given by the tunnel effect through the rectangular
potential barrier
\begin{equation}
V(x)=\left\{
\begin{array}{lr}
0 & x<0 \\
V_{0} & 0<x<a \\
0 & x>a
\end{array}
\right. .
\end{equation}
This problem can be solved exactly and the result\cite{M76} rewritten as
\begin{equation}
T\left( \tilde{E},\lambda \right) =\left\{
\begin{array}{lr}
\frac{4\tilde{E}\left( 1-\tilde{E}\right) }{4\tilde{E}\left( 1-\tilde{E}%
\right) +\sinh ^{2}\left( \sqrt{2\lambda \left( 1-\tilde{E}\right) }\right) }
& 0<\tilde{E}<1 \\
\frac{2}{2+\lambda } & \tilde{E}=1 \\
\frac{4\tilde{E}\left( \tilde{E}-1\right) }{4\tilde{E}\left( \tilde{E}%
-1\right) +\sin ^{2}\left( \sqrt{2\lambda \left( \tilde{E}-1\right) }\right)
} & \tilde{E}>1
\end{array}
\right. ,
\end{equation}
that confirms our prediction that $T$ depends on only two dimensionless
parameters when other strategies produce results in terms of more
dimensional quantities\cite{M76}.

Another interesting and well known example is the Morse oscillator with
potential
\begin{equation}
V(x)=D_{e}\left[ 1-\exp \left( -ax\right) \right] ^{2},  \label{eq:V_Morse}
\end{equation}
where $D_{e},\,a>0$. In this case there are also two obvious possibilities
and we choose $L\equiv 1/a$ so that
\begin{equation}
\tilde{H}=\frac{m}{\hbar ^{2}a^{2}}H=-\frac{1}{2}\frac{d^{2}}{d\tilde{x}^{2}}%
+\lambda \left[ 1-\exp \left( -\tilde{x}\right) \right] ^{2},\;\lambda =%
\frac{mD_{e}}{\hbar ^{2}a^{2}},\;E=\frac{D_{e}}{\lambda }\tilde{E}\left(
\lambda \right) .  \label{eq:H_Morse_dim}
\end{equation}
The bound-state energies of the Morse oscillator are known to be\cite{EWK44}
\begin{eqnarray}
E_{n} &=&hc\left[ \omega _{e}\left( n+1/2\right) -\chi _{e}\omega _{e}\left(
n+1/2\right) ^{2}\right] ,  \nonumber \\
\omega _{e} &=&\frac{a}{\pi c}\sqrt{\frac{D_{e}}{2m}},\;\chi _{e}=\frac{hc}{%
4D_{e}}\omega _{e}.  \label{eq:E_n_Morse}
\end{eqnarray}
that can be easily rewritten as
\begin{equation}
E_{n}=\frac{D_{e}}{\lambda }\left[ \sqrt{2\lambda }\left( n+1/2\right) -%
\frac{1}{2}\left( n+1/2\right) ^{2}\right] ,  \label{eq:E_n_Morse_2}
\end{equation}
in agreement with the prediction of present approach.

In a recent paper Ahmed et al\cite{AKGG19} solved the Schr\"{o}dinger
equation with the potential
\begin{equation}
V(x)=V_{0}\left( 1-e^{2|x|/a}\right) ,\;V_{0}>0,\;a>0,  \label{eq:V_indians}
\end{equation}
that exhibits bound states in the continuum. In their figure 2 they state
``Here, we take $2m/\hbar ^{2}=1$, $V_{0}=50$ and $a=1$.'' Obviously, these
equalities are wrong because the left-hand sides have units and the
right-hand ones do not. If we apply the procedure outlined above we obtain
an equation similar to (\ref{eq:H_f_dim_1}) with $f(\tilde{x})=1-e^{2|\tilde{%
x}|}$ and realize that the model depends on just one parameter $\lambda $
that is dimensionless. Instead of searching for solutions for pairs of
values of $V_{0}$ and $a$ it is sufficient to obtain solutions for just one
parameter $\lambda $. If we calculate $\tilde{E}(\lambda )$ then we have $%
E=V_{0}\tilde{E}(\lambda )$.

Just one more example from a paper published recently. Nguyen and Marsiglio%
\cite{NM20} studied the Schr\"{o}dinger equation with the potential $%
V(x)=-\alpha /x^{2}$ and proposed the alternative truncated potential
\begin{equation}
V_{\epsilon }(x)=\left\{
\begin{array}{c}
-\frac{\alpha }{\epsilon ^{2}}\;\mathrm{if}\;0<x<\epsilon  \\
-\frac{\alpha }{x^{2}}\;\mathrm{if}\;\epsilon <x<\infty
\end{array}
\right. .  \label{eq:V_epsilon}
\end{equation}
If we carry out the change of variables discussed above with $L^{2}=\hbar
^{2}\epsilon ^{2}/(2m\alpha )$ then we obtain the dimensionless Hamiltonian (%
\ref{eq:H_1D_gen_dim}) with the potential
\begin{equation}
\frac{2mL^{2}}{\hbar ^{2}}V_{\epsilon }(L\tilde{x})=\left\{
\begin{array}{c}
-1\;\mathrm{if}\;0<\tilde{x}<\rho _{0} \\
-\frac{\rho _{0}^{2}}{\tilde{x}^{2}}\;\mathrm{if}\;\rho _{0}<\tilde{x}%
<\infty
\end{array}
\right. =\tilde{V}\left( \rho _{0},\tilde{x}\right) ,\;\frac{2mL^{2}}{\hbar
^{2}}=\frac{\epsilon ^{2}}{\alpha },\;\rho _{0}^{2}=\frac{2m\alpha }{\hbar
^{2}}.  \label{eq:V_epsilon_dim}
\end{equation}
Since the dimensionless potential-energy function $\tilde{V}\left( \rho _{0},%
\tilde{x}\right) $ depends on the parameters $\alpha $ and $\epsilon $ only
through the parameter $\rho _{0}$ then the dimensionless energy will depend
only on this parameter: $\tilde{E}\left( \rho _{0}\right) $. Therefore, the
actual energy will be of the form $E=\frac{\alpha }{\epsilon ^{2}}\tilde{E}%
\left( \rho _{0}\right) $. \textit{After} solving the eigenvalue equation in
terms of modified Bessel functions, the authors concluded that $E=-\frac{%
\alpha }{\epsilon ^{2}}\frac{f\left( \rho _{0}^{2}\right) }{\rho _{0}^{2}}$.
Once again, we have been able to predict a general feature of the
quantum-mechanical energies without solving the Schr\"{o}dinger equation.

In closing this section we briefly focus on the time-dependent
Schr\"{o}dinger equation
\begin{equation}
i\hbar \frac{d}{dt}\psi =H\psi ,  \label{eq:Schro_time_dep}
\end{equation}
and proceed as before with respect to the Hamiltonian operator. In addition
to it we define the dimensionless time $\tilde{t}\equiv \omega t$, where $%
\omega $ is an arbitrary frequency. Upon choosing
\begin{equation}
\hbar \omega \equiv \frac{\hbar ^{2}}{mL^{2}},  \label{eq:hb_omega=}
\end{equation}
the Schr\"{o}dinger equation becomes
\begin{equation}
i\frac{d}{d\tilde{t}}\tilde{\psi}=\tilde{H}\tilde{\psi}.
\label{eq:Schro_time_dep_dim}
\end{equation}
In the case of the harmonic oscillator, for example, it follows from
equation (\ref{eq:hb_omega=}) that $\omega =\sqrt{k/m}$ is the oscillator
frequency.

\section{Atoms and molecules}

\label{sec:Atoms_molecules}

The Hamiltonian operator for a system of $K$ particles of masses $m_{i}$,
charges $q_{i}$ at the positions $\mathbf{r}_{i}$, $i=1,2,\ldots ,K$ is
given by\cite{EWK44,P68}
\begin{equation}
H=-\frac{\hbar ^{2}}{2}\sum_{i=1}^{K}\frac{\nabla _{i}^{2}}{m_{i}}%
+\sum_{i=1}^{K-1}\sum_{j=i+1}^{K}\frac{q_{i}q_{j}}{4\pi \epsilon _{0}r_{ij}},
\label{eq:H_atom_molec}
\end{equation}
where $\epsilon _{0}$ is the vacuum permittivity and $r_{ij}=\left| \mathbf{r%
}_{i}-\mathbf{r}_{j}\right| $ is obviously the distance between particles $i$
and $j$. In order to obtain a dimensionless Schr\"{o}dinger equation we
proceed as before and introduce a length unit $L$ and the dimensionless
positions $\tilde{\mathbf{r}}_{i}\equiv \mathbf{r}_{i}/L$ so that $\nabla
_{i}^{2}=L^{-2}\tilde{\nabla}_{i}^{2}$. If $m_{e}$ and $-e$ denote the
electronic mass and charge, respectively, then we define the dimensionless
quantities $\tilde{m}_{i}\equiv m_{i}/m_{e}$ and $\tilde{q}_{i}\equiv
q_{i}/e $ so that the dimensionless Hamiltonian operator $\tilde{H}$ becomes
\begin{equation}
\tilde{H}=\frac{mL^{2}}{\hbar ^{2}}H=-\frac{1}{2}\sum_{i=1}^{K}\frac{\tilde{%
\nabla}_{i}^{2}}{\tilde{m}_{i}}+\frac{m_{e}Le^{2}}{4\pi \epsilon _{0}\hbar
^{2}}\sum_{i=1}^{K-1}\sum_{j=i+1}^{K}\frac{\tilde{q}_{i}\tilde{q}_{j}}{%
\tilde{r}_{ij}}.
\end{equation}
Therefore, if we choose
\begin{equation}
L\equiv \frac{4\pi \epsilon _{0}\hbar ^{2}}{m_{e}e^{2}},
\label{eq:L_atoms_molec}
\end{equation}
the dimensionless Hamiltonian becomes
\begin{equation}
\tilde{H}=-\frac{1}{2}\sum_{i=1}^{K}\frac{\tilde{\nabla}_{i}^{2}}{\tilde{m}%
_{i}}+\sum_{i=1}^{K-1}\sum_{j=i+1}^{K}\frac{\tilde{q}_{i}\tilde{q}_{j}}{%
\tilde{r}_{ij}}.  \label{eq:H_atom_molec_dim}
\end{equation}
It is worth noticing that $L\equiv a_{0}$ is the well known atomic unit of
length and
\begin{equation}
\frac{\hbar ^{2}}{m_{e}a_{0}^{2}}=\frac{e^{2}}{4\pi \epsilon _{0}a_{0}},
\label{eq:atomic_unit_energy}
\end{equation}
is the atomic unit of energy. This equation also shows that $a_{0}$ makes a
term that looks as a kinetic energy (left) equal to other term that looks
like a potential (right). As argued above, one of the most noticeable
advantages of this procedure is that we get rid of small numbers like $\hbar
$, $e$, $m_{e}$, $\epsilon _{0}$, etc. It is equivalent to setting these
quantities equal to unity. Such small numbers may increase the round-off
errors in a numerical calculation of atomic and molecular properties and for
this reason atomic units are used throughout\cite{EWK44,P68}.

The potential-energy function of this system of particles is invariant under
space translations and, consequently, we should remove the free motion of
the center of mass before applying any approximate method to the
Schr\"{o}dinger equation\cite{F08b,FE10} (and references therein). However,
this issue is not relevant to present discussion because we do not solve any
equation here.

As an example, consider the Hamiltonian operator for the hydrogen atom\cite
{EWK44,P68}
\begin{equation}
H=-\frac{\hbar ^{2}}{2m}\nabla ^{2}-\frac{e^{2}}{4\pi \epsilon _{0}r},
\label{eq:H_hydrogen}
\end{equation}
where $m\equiv m_{e}m_{n}/\left( m_{e}+m_{n}\right) $ is the reduced mass of
the system, $m_{n}$ the nuclear mass and $r$ the distance between the
nucleus and the electron. The reduced mass appears when we remove the motion
of the center of mass as mentioned above. In atomic units the Hamiltonian (%
\ref{eq:H_hydrogen}) becomes
\begin{equation}
\tilde{H}=-\frac{1}{2\tilde{m}}\tilde{\nabla}^{2}-\frac{1}{\tilde{r}},\;%
\tilde{m}=\frac{m_{n}}{m_{n}+m_{e}}.  \label{eq:H_hydrogen_dim}
\end{equation}
If instead of the unit of length (\ref{eq:L_atoms_molec}) we choose
\begin{equation}
L\equiv \frac{4\pi \epsilon _{0}\hbar ^{2}}{me^{2}},
\end{equation}
the Hamiltonian operator for hydrogen takes an even simpler form
\begin{equation}
\tilde{H}=-\frac{1}{2}\tilde{\nabla}^{2}-\frac{1}{\tilde{r}}.
\label{eq:H_hidrogen_dim_2}
\end{equation}

In the case of atoms it is common usage to resort to the so called
clamped-nucleus approximation, which for hydrogen can be expressed as
follows:
\begin{equation}
\lim\limits_{m_{n}\rightarrow \infty }\tilde{m}=1.
\end{equation}
Within this approximation it is not necessary to remove the motion of the
center of mass because it is located at the nucleus that remains fixed at
origin\cite{EWK44,P68}. However, in the case of highly accurate calculations
(which may also include relativistic effects) the mass-polarization terms
due to the nuclear motion should be taken into consideration.

\section{Perturbation theory}

\label{sec:PT}

In this section we show that suitable dimensionless Schr\"{o}dinger
equations may facilitate the application of perturbation theory\cite{F01}.
Since, as already pointed out above, we do not solve the Schr\"{o}dinger
equation in this paper we will not be concerned with the convergence
properties of the perturbation series.

The first example is the widely discussed quartic anharmonic oscillator that
we write in the following way
\begin{equation}
H=-\frac{\hbar ^{2}}{2m}\frac{d^{2}}{dx^{2}}+V(x),\;V(x)=\frac{k_{2}}{2}%
x^{2}+k_{4}x^{4},\;k_{2},k_{4}>0.  \label{eq:H_QAO}
\end{equation}
On applying the strategy outlined in section~\ref{sec:one-dim} we have
\begin{equation}
\frac{mL^{2}}{\hbar ^{2}}V\left( L\tilde{x}\right) =\frac{mk_{2}L^{4}}{%
2\hbar ^{2}}\tilde{x}^{2}+\frac{mk_{4}L^{6}}{\hbar ^{2}}\tilde{x}^{4}.
\end{equation}
In this case we can try two choices of the length unit $L$, the first one is
the harmonic oscillator length $L\equiv \left[ \hbar ^{2}/(mk_{2})\right]
^{1/4}$ that leads to
\begin{eqnarray}
\tilde{H} &=&\frac{H}{\hbar \omega }=-\frac{1}{2}\frac{d^{2}}{d\tilde{x}^{2}}%
+\frac{\tilde{x}^{2}}{2}+\lambda \tilde{x}^{4},  \nonumber \\
\omega &=&\sqrt{\frac{k_{2}}{m}},\;\lambda =\frac{\hbar k_{4}}{\left(
mk_{2}^{3}\right) ^{1/2}}=\frac{\hbar k_{4}}{m^{2}\omega ^{3}},  \nonumber \\
\tilde{E}(\lambda ) &=&\frac{E}{\hbar \omega }.  \label{eq:H_QAO_dim_1}
\end{eqnarray}
If we apply perturbation theory we obtain the $\lambda $-power series
\begin{equation}
E=\hbar \omega \sum_{j=0}^{\infty }\tilde{E}^{(j)}\lambda ^{j},
\end{equation}
that is suitable for sufficiently small values of $\lambda $. There are
several efficient methods for the calculation of the coefficients $\tilde{E}%
^{(j)}$ in exact analytical form to any desired order $j$\cite{F01}.

A second choice is
\begin{equation}
L\equiv \left( \frac{\hbar }{\sqrt{mk_{4}}}\right) ^{1/3}=\left( \frac{\hbar
}{m\omega \lambda ^{1/3}}\right) ^{1/2},
\end{equation}
that leads to
\begin{eqnarray}
\tilde{H} &=&\frac{H}{\hbar \omega \lambda ^{1/3}}=-\frac{1}{2}\frac{d^{2}}{d%
\tilde{x}^{2}}+\frac{\tilde{x}^{2}}{2\lambda ^{2/3}}+\tilde{x}^{4},
\nonumber \\
\tilde{E}(\lambda ) &=&\frac{E}{\hbar \omega \lambda ^{1/3}}.
\label{eq:H_QAO_dim_2}
\end{eqnarray}
This equation suggests that we can expand the energies as
\begin{equation}
E=\hbar \omega \lambda ^{1/3}\sum_{j=0}^{\infty }\tilde{e}^{(j)}\lambda
^{-2j/3}.
\end{equation}
In this case we cannot obtain the expansion coefficients $\tilde{e}^{(j)}$
exactly but the mere knowledge of the existence of this series is useful in
the application of some resummation methods\cite{F01}. It is worth pointing
out that a calculation for a single value of $\lambda $ is equivalent to an
infinite number of calculations based on variations of $m$, $k_{2}$ and $%
k_{4}$ such that $k_{4}/\left( mk_{2}^{3}\right) ^{1/2}$ is constant. This
fact is an obviously useful advantage of resorting to a dimensionless
equation.

Let us now consider a one-dimensional Hamiltonian operator with the
potential (\ref{eq:V=V0_f}), where $V_{0}>0$ and $f(q)$ exhibits a minimum
at $q=0$ such that $f(0)=0$. We assume that $f(q)$ can be expanded in a
Taylor series
\begin{equation}
f(q)=\sum_{j=2}^{\infty }\frac{f_{j}}{j!}q^{j}.  \label{eq:f(q)_Taylor}
\end{equation}
The dimensionless Hamiltonian reads
\begin{equation}
\tilde{H}=\frac{mL^{2}}{\hbar ^{2}}H=-\frac{1}{2}\frac{d^{2}}{d\tilde{x}^{2}}%
+\frac{mL^{4}V_{0}f_{2}}{2\hbar ^{2}a^{2}}\tilde{x}^{2}+\sum_{j=3}^{\infty }%
\frac{mV_{0}L^{j+2}}{j!\hbar ^{2}a^{j}}f_{j}\tilde{x}^{j}.
\end{equation}
If we choose
\begin{equation}
L\equiv \left( \frac{\hbar ^{2}a^{2}}{mf_{2}V_{0}}\right) ^{1/4},
\end{equation}
then the dimensionless Hamiltonian becomes
\begin{equation}
\tilde{H}=-\frac{1}{2}\frac{d^{2}}{d\tilde{x}^{2}}+\frac{1}{2}\tilde{x}%
^{2}+\sum_{j=1}^{\infty }\frac{f_{j+2}}{f_{2}}\lambda ^{j}\tilde{x}%
^{j+2},\;\lambda \equiv \frac{L}{a}=\left( \frac{\hbar ^{2}}{ma^{2}f_{2}V_{0}%
}\right) ^{1/4},
\end{equation}
which shows that the energies can be expanded as
\begin{equation}
E=\hbar \sqrt{\frac{V_{0}f_{2}}{ma^{2}}}\sum_{j=0}^{\infty }\tilde{E}%
^{(j)}\lambda ^{j}.
\end{equation}
There are efficient approaches for the exact analytical calculation of the
coefficients $\tilde{E}^{(j)}$\cite{F01}.

The last example is the Hamiltonian operator for an atom with $N$ electrons
and nuclear charge $Ze$ in the clamped-nucleus approximation
\begin{eqnarray}
H &=&H_{0}+H^{\prime },  \nonumber \\
H_{0} &=&-\frac{\hbar ^{2}}{2m_{e}}\sum_{i=1}^{N}\nabla
_{i}^{2}-\sum_{i=1}^{N}\frac{Ze^{2}}{4\pi \epsilon _{0}r_{i}},  \nonumber \\
H^{\prime } &=&\sum_{i=1}^{N-1}\sum_{j=i+1}^{N}\frac{e^{2}}{4\pi \epsilon
_{0}r_{ij}},  \label{eq:H_atom}
\end{eqnarray}
where $r_{i}$ is the distance between the electron $i$ and the nucleus and $%
r_{ij}$ the distance between a pair of electrons. Since the Schr\"{o}dinger
equation for $H_{0}$ is exactly solvable we can apply perturbation theory
where $H^{\prime }$ is the perturbation. In what follows we show that the
dimensionless equation gives us valuable information about the solution
derived from perturbation theory.

As in the previous examples, the dimensionless Hamiltonian is
\begin{equation}
\tilde{H}=\frac{m_{e}L^{2}}{\hbar ^{2}}H=-\frac{1}{2}\sum_{i=1}^{N}\tilde{%
\nabla}_{i}^{2}-\sum_{i=1}^{N}\frac{m_{e}LZe^{2}}{4\pi \epsilon _{0}\hbar
^{2}\tilde{r}_{i}}+\sum_{i=1}^{N-1}\sum_{j=i+1}^{N}\frac{m_{e}Le^{2}}{4\pi
\epsilon _{0}\hbar ^{2}\tilde{r}_{ij}}.
\end{equation}
In this case we choose
\begin{equation}
L\equiv \frac{4\pi \epsilon _{0}\hbar ^{2}}{m_{e}Ze^{2}}=\frac{a_{0}}{Z},
\label{eq:L_atom}
\end{equation}
and obtain
\begin{equation}
\tilde{H}=-\frac{1}{2}\sum_{i=1}^{N}\tilde{\nabla}_{i}^{2}-\sum_{i=1}^{N}%
\frac{1}{\tilde{r}_{i}}+\frac{1}{Z}\sum_{i=1}^{N-1}\sum_{j=i+1}^{N}\frac{1}{%
\tilde{r}_{ij}}.  \label{eq:H_atom_dim}
\end{equation}
It is clear that the result of the application of perturbation theory as
indicated above will be a series of the form
\begin{equation}
E=\frac{\hbar ^{2}Z^{2}}{m_{e}a_{0}^{2}}\sum_{j=0}^{\infty }\tilde{E}%
^{(j)}Z^{-j}.  \label{eq:E_atom_series}
\end{equation}
Since the perturbation coefficients $\tilde{E}^{(j)}$ only depend on $N$,
then if the number of electrons remains constant we conclude that the rate
of convergence of this perturbation series will improve with $Z$.

\section{Conclusions}

\label{sec:conclusions}

This paper shows the advantages of using dimensionless equations in
non-relativistic quantum mechanics. The dimensionless Schr\"{o}dinger
equation is simpler than the original one which facilitates the process of
obtaining the desired solutions. If one has to resort to a numerical method
the dimensionless equation (with all its quantities of the order of unity)
is considerably less prone to round-off errors. When we derive a
dimensionless equation we realize which are the relevant parameters that
should appear in the solution beforehand. In the case of perturbation theory
we can predict the general form of the solution and obtain a suitable
perturbation parameter. It is also important to realize that the proper
scaling of the variables is by no means guesswork. A suitable definition for
$L$ is dictated by the form of the equation for the physical problem. We
believe that it is worthwhile to teach this approach in undergraduate as
well as graduate courses on quantum mechanics. It is worth adding that
dimensionless equations are also useful in other areas of physics, such as,
for example, classical physics\cite{AF18,F18}.

\end{document}